\documentclass{article}

\usepackage{arxiv}

\usepackage[utf8]{inputenc} 
\usepackage[T1]{fontenc}    
\usepackage{hyperref}       
\usepackage{url}            
\usepackage{booktabs}       
\usepackage{amsfonts}       
\usepackage{nicefrac}       
\usepackage{microtype}      
\usepackage{lipsum}		
\usepackage{graphicx}
\usepackage{natbib}
\usepackage{doi}
\usepackage{soul}
\usepackage{tikz}
\usepackage{xcolor}
\usepackage{booktabs}
\definecolor{gray92}{gray}{0.92}
\definecolor{black}{RGB}{0,0,0}
\definecolor{FD8D3C}{RGB}{253,141,60}
\definecolor{A1D99B}{RGB}{161,217,155}
\definecolor{74C476}{RGB}{116,196,118}
\definecolor{41AB5D}{RGB}{65,171,93}
\definecolor{F768A1}{RGB}{247,104,161}
\definecolor{238B45}{RGB}{35,139,69}
\definecolor{005A32}{RGB}{0,90,50}
\definecolor{00441B}{RGB}{0,68,27}
\definecolor{6E016B}{RGB}{110,1,107}
\definecolor{4292C6}{RGB}{66,146,198}
\definecolor{8C6BB1}{RGB}{140,107,177}
\definecolor{gray85}{gray}{0.85}
\definecolor{gray10}{gray}{0.10}
\definecolor{gray30}{gray}{0.30}
\definecolor{gray20}{gray}{0.20}
\definecolor{gray80}{gray}{0.80}
\definecolor{1F78B4}{RGB}{31,120,180}
\definecolor{FDBF6F}{RGB}{253,191,111}

\usepackage{amsmath,amsfonts,amsthm,amssymb,thmtools}
\usepackage{bm}
\usepackage{float}
\usepackage{subcaption}

\usepackage{soul,color}

\newcommand{\bxi}{{\bm{x}_i}}
\newcommand{\bx}{{\bm{x}}}
\newcommand{\bXi}{{\bm{X}_i}}
\newcommand{\bX}{{\bm{X}}}

\newcommand{\tacwt}{{\hat{\tau}^{ACW-t}}}
\newcommand{\tadj}{{\hat{\tau}^{SS}}}
\newcommand{\taipw}{{\hat{\tau}^{fIPW}}}
\newcommand{\tcw}{{\hat{\tau}^{CW}}}
\newcommand{\treg}{{\hat{\tau}^{OLS}}}
\newcommand{\tregrct}{{\hat{\tau}^{OLS}(\bX_{\{i\in \mathcal{R}\}})}}
\newcommand{\tregrctf}{{\hat{\tau}^{OLS}(\bx_{\{i\in \mathcal{R}\}})}}
\newcommand{\tregy}{{\hat{\tau}^{OLS}(\yhatn)}}
\newcommand{\tregyaux}{{\hat{\tau}^{OLS}(\yhat)}}
\newcommand{\tdm}{{\hat{\tau}^{DM}}}

\newcommand{\yhati}{{\hat{Y}_i(\cdot)}}
\newcommand{\yhat}{{\hat{\bm Y}_{\mathcal{O}}(\cdot)}}
\newcommand{\yhatn}{{\hat{\bm Y}_{\mathcal{O}}}}
\newcommand{\yhato}{{\hat{\bm Y}_{\mathcal{O}}}}

\newcommand{\gstar}{{\bm{G}^{\ast}}}
\newcommand{\g}{{\bm{G}}}
\newcommand{\syn}{{\tilde{\bm{D}}}}

\newcommand{\V}{\mathbb{V}}

\title{Combining observational and experimental data for causal inference considering data privacy}

\date{July, 2024}	

\author{ Charlotte Z.~Mann \\
Department of Statistics\\
University of Michigan\\
Ann Arbor, MI, USA\\
\texttt{manncz@umich.edu}
	\And
	Adam C. Sales\\
 Department of Mathematical Sciences\\
	Worcester Polytechnic Institute\\
 Worcester, MA, USA\\
 \texttt{asales@wpi.edu} \\
 \And 
Johann A.~Gagnon-Bartsch\\
Department of Statistics\\
University of Michigan\\
Ann Arbor, MI, USA\\
\texttt{johanngb@umich.edu}
}

\renewcommand{\headeright}{}
\renewcommand{\undertitle}{}
\renewcommand{\shorttitle}{Data integration with disclosure limitation}

\begin{document}
\maketitle

\begin{abstract}
	

Combining observational and experimental data for causal inference can improve treatment effect estimation. However, many observational data sets cannot be released due to data privacy considerations, so one researcher may not have access to both experimental and observational data. Nonetheless, a small amount of risk of disclosing sensitive information might be tolerable to organizations that house confidential data. In these cases, organizations can employ data privacy techniques, which decrease disclosure risk, potentially at the expense of data utility. In this paper, we explore disclosure limiting transformations of observational data, which can be combined with experimental data to estimate the sample and population average treatment effects. We consider leveraging observational data to improve generalizability of treatment effect estimates when a randomized experiment (RCT) is not representative of the population of interest, and to increase precision of treatment effect estimates. Through simulation studies, we illustrate the trade-off between privacy and utility when employing different disclosure limiting transformations. We find that leveraging transformed observational data in treatment effect estimation can still improve estimation over only using data from an RCT.
\end{abstract}

\section{Introduction}\label{sec:intro}

A growing literature is developing ways to combine experimental and observational studies for causal inferences \citep{colnet_causal_2024}. While treatment effect estimates from randomized experiments (RCTs) can be free from confounding bias, observational studies generally provide a richer source of information on a population of interest. The internet has given rise to more observational datasets that may be used to address the same questions as randomized experiments. Therefore, there should be more opportunities to leverage observational and RCT data together for improved treatment effect estimation. However, in practice it is not always the case that a researcher has access to both sources of data due to data privacy \cite{snoke_how_2020}. 

Many government agencies have useful data that they cannot release to the public in order to preserve data privacy. Data privacy refers to the right of individuals whom the data describe to control what information about themselves is shared \cite{fellegi_question_1972, raghunathan_synthetic_2021}. Typically sensitive data that is released to the public is sanitized in various ways, which potentially render the released data less useful. For example, only aggregate statistics or a sample of the data may be released, or small values are censored. There is a trade-off to consider, between data privacy, a right to be upheld, and the amount of information researchers can access to answer societal questions.
 
Balancing data privacy and releasing useful information is ultimately a policy decision. There are currently no overarching policies in the U.S. regulating data privacy or confidentiality \cite{schwartz_eu-us_2013, bellovin_privacy_2018}. Rather, data privacy is legislated on a sector-by-sector basis \cite{schwartz_eu-us_2013}. For example, patient, student, financial, and additional data from U.S. governmental agencies data are protected by separate acts: HIPPA, FERPA, FCRA, and CIPSEA (Health Insurance Portability and Accountability Act; Family Educational Rights and Privacy Act; Fair Credit Reporting Act; Confidential Information Protection and Statistical Efficiency Act) \cite{schwartz_eu-us_2013, wood_differential_2018}. On the other hand, members of the European Union and other countries throughout the world have adopted data privacy legislation with broad scopes across sectors \cite{schwartz_eu-us_2013}. We do not address the legality of different approaches to data privacy. Rather, we raise these issues to establish that data stewards operate in different legal contexts, which may or may not provide specific procedures for protecting data privacy. 

We consider the setting in which analysts of an RCT could potentially use auxiliary, observational data to improve treatment effect estimation through data integration. However, the relevant auxiliary data cannot be released in its raw form. We aim to address the primary research question: Can privacy-preserving releases of confidential observational data be used to improve causal estimation when integrated with the RCT data?

There are two broad ways that integrating observational data into and experimental treatment effect estimates is often used to improve causal estimation, (1) estimating treatment effects for a population of interest when the RCT sample is not representative of that population and (2) increasing precision of RCT estimates. We consider two previously-developed data integration methods for causal inference, which each address one of these aims. These methods are are well-suited for our investigation because they only require summaries of the auxiliary, observational data. Then, we consider ways that stewards of such observational data could transform and release the data to preserve data privacy, as discussed in the data privacy literature. We focus on transformations that and can still be used in data integration methods for treatment effect estimation. Finally, through simulation studies, we illustrate the trade-off between privacy and utility when integrating different releases of the private, observational data in RCT treatment effect estimation, both to generalize to populations of interest, and to improve experimental precision.

The work in this paper is distinct from previous work on releasing privacy-protected causal estimates, such as in \cite{kusner_private_2016, wang_towards_2020, ma_noleaks_2022}, in a couple of ways. First, we do not consider releasing a private causal \textit{estimate}, but rather how a private data release itself could be used in causal estimation with data integration. Second, \cite{kusner_private_2016, wang_towards_2020, ma_noleaks_2022} rely on a different causal framework than we do in this paper.

We aim for this work to inform data stewards who want to release data that is as useful as possible while balancing data privacy. We also aim to encourage conversation between the literature that combines experimental and observational data for causal inference and the data privacy literature.

This paper is organized as follows. Section~\ref{sec:notation} establishes notation and the causal estimands and estimators of interest. Section~\ref{sec:priv} provides background on data privacy techniques and presents disclosure limiting transformations of auxiliary data. Section~\ref{sec:sims} describes simulation studies to evaluate the utility of the proposed transformations in treatment effect estimation that integrates transformed auxiliary data and RCT data and discusses the results. Section~\ref{sec:disc} concludes.

\section{Leveraging Observational and Experimental Data in Causal Inference}\label{sec:notation}

The presentation in this section primarily follows \cite{colnet_causal_2024}. Consider a randomized experiment with $n$ subjects and related auxiliary, observational data with $m$ subjects, both sampled from some population. For example, in a medical trial conducted to evaluate a treatment for sleep apnea there may also be electronic health records for thousands of patients across a country with sleep apnea. We denote $\mathcal{R} = \{i: 1, ..., n\}$ as the index set for subjects in the randomized experiment and $\mathcal{O} = \{i: n+1, ..., m+n\}$ as the index set for subjects in the auxiliary (observational) data. Let $S_i$ be an indicator of whether subject $i$ is in the RCT sample, and $\pi_i^S = P(S_i = 1)$ be the probability of selection into the RCT. The probability of selection into the RCT may depend on subject $i$'s observable characteristics, so that the RCT sample is not necessarily representative of the population.  Let $T$ be a binary treatment indicator so $T_i = 1$ if subject $i$ is assigned to treatment and $T_i = 0$ if subject $i$ is assigned to control. We assume the probability that a subject in the RCT is assigned treatment, $\pi_i = P(T_i = 1), 0 < \pi_i < 1$, is known.  In the setting we consider, we do not require the treatment to be observed in the observational data. For each subject in the RCT and observational samples, we observe a vector of covariates $\bm X_i$. Throughout this paper, the following conventions will be used: fixed quantities are in lowercase, random quantities are in uppercase, scalars are non-bolded, and vectors or matrices are bolded.

Following Neyman \cite{neyman:1935} and Rubin \cite{rubin_estimating_1974}, each subject has two potential outcomes, one which would be observed if treated, $Y^t_i$, and the other under control, $Y^c_i$. The observed outcome is a function of the potential outcomes and the treatment assignment: $Y_i = T_iY^t_i + (1-T_i)Y^c_i.$ We assume no interference, i.e. $Y_i \perp \{T_j, j\neq i\}$ \citep{angrist_identification_1996}.

Researchers may be interested in a number of causal estimands. This work explores integrating experimental and observational data, under data privacy considerations, to estimate the the population average treatment effect (PATE), as well as the RCT sample average treatment effect (SATE). The PATE is the expected treatment effect for a population of interest: $\tau_{PATE} = \mathbb{E}[Y^t - Y^c]$. The SATE is the expected treatment effect for an RCT sample: $\tau_{SATE} =  \mathbb{E}[Y^t - Y^c| S = 1]$. Combining experimental and observational data can improve estimating both the PATE and the SATE in different ways, as discussed in the following sections.

\subsection{Generalizing to Populations of Interest}\label{sec:gen_est}

Treatment effects estimated only with an RCT do not necessarily generalize to a population of interest. Consider estimating the PATE for the population from which subjects of an RCT were sampled. If the subjects in the RCT are systematically different from this population (i.e., there is some selection bias to inclusion in the RCT), and the treatment effect depends on subject characteristics, then an estimate of the PATE with only the RCT data is biased. Further it will be unclear how ``far off'' the estimate is. In this case, we say that the RCT estimate does not generalize to the population. Observational data is often more representative of the population from which they were sampled. Therefore, to estimate the PATE, it is useful to leverage information from auxiliary observational data.
 
The literature proposes a number of estimates for the PATE which integrate 
 experimental and observational data \citep{cole_generalizing_2010, stuart_use_2011, lesko_generalizing_2017, dahabreh_generalizing_2019, lee_improving_2021}. See \cite{colnet_causal_2024} for a full review. We focus on the calibration weighted (CW) estimator proposed by \cite{lee_improving_2021}, for two primary reasons. First, a statistical summary of auxiliary data is sufficient for the CW estimator. As will be discussed in detail in Section~\ref{sec:priv}, statistical summaries of data already do some of the work to limit disclosure of confidential data. Thus, the CW estimator is better-suited to take private releases of auxiliary data as an input than other estimators, which rely on the full auxiliary data, at the subject level. Second, the results of the review in \cite{colnet_causal_2024} indicate that the CW estimator outperforms other options across different settings.
 
The CW estimator \cite{lee_improving_2021} is a variation of the inverse probability weighted (IPW) estimator \citep{horvitz_generalization_1952, robins_estimation_1994}, $$ \hat{\tau}^{IPW} = \sum_{i \in \mathcal{R}}\Big[T_i\frac{Y_i}{\pi_i} - (1-T_i)\frac{Y_i}{1-\pi_i}\Big].$$ The IPW estimator only uses data from the RCT ($i \in \mathcal{R})$ so it may be biased for the PATE if there is selection bias. The CW estimator re-weights the RCT subjects in the IPW estimator. The goal of the weighting is to align the empirical covariate distribution of the RCT subjects with that of the the auxiliary subjects ($i \in \mathcal{O})$, which are assumed to represent the population of interest. In addition to the assumptions discussed previously, the CW estimator relies on the following assumptions to identify the PATE: (1) positivity of selection into the RCT ($0 < \pi_i^S < 1$), and (2) that the conditional average effect (CATE) for the RCT sample is equivalent to the population CATE ($\mathbb{E}[Y^t - Y^c| \bm X, S=1] = \mathbb{E}[Y^t - Y^c| \bm X]$). 

The CW estimator is defined as: $$ \hat{\tau}^{CW} = \sum_{i \in \mathcal{R}}\hat{w}(\bXi)\Big[T_i\frac{Y_i}{\pi_i} - (1-T_i)\frac{Y_i}{1-\pi_i}\Big]$$
Each subject $i \in \mathcal{R}$ is assigned a weight $\hat{w}(\cdot)$, which is estimated using covariates from the auxiliary study and the RCT, solving the optimization problem: $\underset{w_1, ..., w_n}{\min} \sum_{i=1}^n w_i \log w_i$, subject to $w_i \geq 0$, $\sum_{i=1}^n w_i = 1$ and 
 
\begin{equation}\label{eq:calib}
    \sum_{i \in \mathcal{R}} w_i g(\bXi)  = \frac{1}{m} \sum_{i \in \mathcal{O}}g(\bXi).
\end{equation}

Equation~\ref{eq:calib} is the key restriction on $w_i$ for generalizability - the goal is for the weighted sum of $g(\bXi)$ in the RCT sample to be equal to the simple mean of $g(\bXi)$ in the auxiliary sample. Two reasonable choices for $g(\cdot)$ would be $g(\bm z) = \bm z$ and $g(\bm z) = \bm z \bm z'$. With these choices, the first or second empirical moments of the RCT sample covariates are calibrated to those moments of the auxiliary sample covariates since $\frac{1}{m} \sum_{i \in \mathcal{O}}g(\bXi)$ is a consistent estimator of $E[g(\bm X)]$ for the auxiliary study. \cite{lee_improving_2021} note that the calibration weighting estimator is a consistent estimator for the PATE if either (1) the probability of RCT participation can be modeled as $\exp\{\eta_0^Tg(\bm X)\}$ for some $\eta_0$ or (2) the CATE is a linear function of $g(\bm X)$.

\subsection{Improving Experimental Precision}\label{sec:prec_est}

We next consider a context in which the treatment effect for only an RCT sample is of interest. Even though there may no longer be a larger population of interest, incorporating information from observational data can still prove useful, in this case by improving the precision of an estimate of the SATE. 

RCTs often have small sample sizes, so RCT estimates of the SATE may lack precision. A common approach to improving precision is covariate adjustment. Covariate adjustment accounts for variance in $Y_i$ which is not due to the treatment, but rather some observed covariates. A popular adjusted estimator for the SATE is the estimated coefficient for the treatment assignment in a regression model. One might consider a linear model $Y_i = \alpha + \tau T_i + \bm \beta \bXi +\varepsilon_i$ with the standard OLS assumptions. We denote the estimate of the coefficient on $T_i$ in this model, using OLS and the RCT sample only, as $\tregrct$. The increase in precision achieved by covariate adjustment depends on how much of the variance in $Y_i$ can be explained by the observed covariates. 

Because observational studies typically have much larger sample sizes than randomized experiments ($m >> n$), including information from auxiliary observational data can improve precision more than covariate adjustment with the RCT sample alone. There are various ways that auxiliary information can be integrated into RCT analysis to improve precision \citep{pocock_combination_1976, viele_use_2014, aronow_class_2013, deng_improving_2013, sales_rebar_2018,gui_combining_2020, opper_improving_2021, gagnon-bartsch_precise_2023}. For example, one part of the literature leverages historical controls from observational studies or previously run RCTs to improve precision in RCT estimates by pooling the different sources of data \citep{pocock_combination_1976, viele_use_2014}. We focus on an approach that uses auxiliary observational data to construct a highly predictive covariate for the outcome of interest in the RCT as in \citep{aronow_class_2013, sales_rebar_2018, opper_improving_2021,  gagnon-bartsch_precise_2023}. We will call the general method the ``super-covariate data integration approach.''

The first step of the super-covariate data integration approach is to fit a model of the outcome of interest (Y)  with the \textit{auxiliary data only}. The only goal of this model is to predict the outcome of interest as well as possible in the RCT sample. The only requirements for the auxiliary prediction model are that (1) it uses only covariates that would be considered pre-treatment or baseline in the RCT and (2) predictions can be generated for the RCT sample (i.e. the same baseline covariates need to be available in the RCT and auxiliary data). Let $\hat{y}_{\mathcal{O}}(\cdot)$ denote such an auxiliary model. The second step of the approach is to generate outcome predictions for the RCT sample. let $\hat{Y}_i = \hat{y}_{\mathcal{O}}(\bm X_i)$ denote a prediction for subject $i$ from the model fit on the auxiliary data and $\yhato$ denote the $n \times 1$ vector of such predictions for the RCT subjects ($i \in \mathcal{R}$). Then, $\yhato$ can be considered a baseline covariate because it is independent of the treatment assignment in the RCT sample. As the final step of the super-covariate approach approach, we replace the RCT covariates $\bm X_{\{i \in \mathcal{R}\}}$ with $\yhato$ in any covariate-adjusted causal estimator. In this paper, we will replace the covariates in the regression estimator (i.e. estimate $\tau$ with OLS in the model $Y_i = \alpha + \tau T_i + \beta \hat{Y}_i +\varepsilon_i$). We denote this regression estimator, using the auxiliary prediction as a covariate as $\tregy$.

Especially when there are a large number of covariates, a model fit on (large) auxiliary data can be more informative of the outcome of interest than a model fit on (small) RCT data. For this reason, we expect $\yhato$ to be a powerful covariate to adjust for the variability in the RCT outcome, which is not explained by the treatment assignment (i.e. a ``super-covariate''). Thus, adjusting with the auxiliary predictions can improve precision beyond covariate adjustment with the RCT covariates \cite{sales_rebar_2018, opper_improving_2021,  gagnon-bartsch_precise_2023}. 

We focus on this data integration approach for the SATE because it is highly flexible, relies on few assumptions, and is highly efficacious. All that is required from the auxiliary data is a model of the outcome of interest. Thus, as with the CW estimator, only a statistical summary of the auxiliary data is required. Additionally, this method relies on very few assumptions with regards to the auxiliary and RCT data. The auxiliary model need not be ``correct'' in any sense. The only requirement for the approach to improve precision is that the auxiliary model predicts the outcomes in the RCT setting well. Finally, there is a body of previous literature supporting the efficacy of this approach in general \citep{sales_rebar_2018, opper_improving_2021, sales_more_2022, gagnon-bartsch_precise_2023}, so we are interested in whether that efficacy can be maintained even if using privacy-preserving transformations of auxiliary data, rather than the original data itself.

\section{Disclosure Limiting Transformations of Auxiliary Data}\label{sec:priv}

In the previous section we discussed two estimators that combine experimental and observational data. In practice, one entity may not have access to both types of data. We assume that analysts with access to a randomized experiment are interested in incorporating information from a relevant auxiliary, observational, study. However, the data stewards of the auxiliary study cannot release the data ($(Y_i, \bXi), i \in \mathcal{O}$) due to data privacy. Therefore, we consider transformations of the restricted auxiliary data that limit disclosure risk and can be used in the estimators discussed in the previous section ($\tcw$ and $\tregy$). In this section we introduce data privacy frameworks (\ref{sec:priv-exp}), discuss the properties of differentially private algorithms (\ref{sec:priv-dp}), and define the disclosure limiting transformations that we compare in this work (\ref{sec:priv-transforms}).

\subsection{Data Privacy Overview}\label{sec:priv-exp}

With the rise of the internet and an explosion of data availability, computer scientists and statisticians have been considering issues of data privacy for the past four decades \citep[see][for reviews]{duncan_enhancing_1991, fienberg_invited_2000, aggarwal_privacy-preserving_2008, matthews_data_2011, salas_basics_2018, slavkovic_statistical_2022}. Two distinct but related concepts are data privacy and data confidentiality. Data privacy is defined as the right of individuals to control information about themselves \cite{fellegi_question_1972}. Data confidentiality is the agreement between individuals and data stewards regarding the extent to which others can access any private/sensitive information provided \cite{fellegi_question_1972, raghunathan_synthetic_2021}. Disclosure risk refers to the risk that an attacker could access sensitive information from released data. Historically there was a focus on microdata, data which has information at an individual level and includes information that could be used to identify an individual, however more recent work considers the risk of disclosure for statistical summaries \citep{slavkovic_statistical_2022}. The goal of data privacy methods is to reduce disclosure risk.

We can view data privacy techniques as falling into two primary frameworks: statistical disclosure control (SDC) and differential privacy (DP) \cite[see][for detailed discussions]{slavkovic_statistical_2022, raghunathan_synthetic_2021}. SDC aims to uphold data privacy by maintaining data confidentiality. SDC techniques include, for example, synthetic data, cell suppression, data swapping, matrix masking and noise additions to limit the risk for disclosure of individual identities and attributes \cite{matthews_data_2011}. There is not one measure of disclosure risk in the SDC framework. On the other hand, the DP framework provides a mathematical definition for disclosure risk, although for a specific type of risk (discussed in the following section). We consider techniques under both frameworks in this paper, and use the term disclosure limitation to mean limiting the risk of disclosing sensitive information, not specifically referring to SDC.

With any disclosure limiting procedure, there is a trade-off between privacy and utility \cite{matthews_data_2011, bowen_personal_2021, slavkovic_statistical_2022}. To maximize utility, analysts would want the original data, and thus there would be no data privacy. On the other hand, for there to be no disclosure risk, the data could not be released, so there would be no utility. Data stewards therefore must decide on a tolerable disclosure risk and a measure of data utility in order to balance the two competing forces. This is not a trivial task. As discussed in Section~\ref{sec:intro}, depending on the context, there are not necessarily formal guidelines for determining a tolerable disclosure risk. Additionally, anticipating all of the desired uses of a dataset is not feasible. In this paper, we focus on specific uses for the data releases, which are associated with specific metrics to assess utility. We do not specify a tolerable disclosure risk. Rather, the simulation studies in the Section~\ref{sec:sims} explore the privacy-utility trade-off when employing different disclosure limiting transformations.

\subsection{Differential Privacy}\label{sec:priv-dp}

\cite{dwork_calibrating_2006, dwork_differential_2006} introduced differential privacy in the early 2000's. Differential privacy is considered the first rigorous mathematical quantification of disclosure risk for privacy-preserving algorithms. Differentially private (DP) algorithms limit the difference in distribution between outputs of the algorithm generated from datasets that differ by only one observation \cite{bowen_philosophy_2021}. 

Consider a private dataset $\bm d$ and a dataset $\bm d'$ which is a subset of $\bm d$, differing by one observation: $d(\bm d, \bm d') = 1$.\footnote{$d(\bm d, \bm d') = 1$ has two different definitions in the differential privacy literature. It can either refer to removing one observation or to changing all of the values of one observation \cite{barrientos_feasibility_2023}. We will assume $d(\bm d, \bm d') = 1$ means that one observation is removed in this paper.} Formally, a random algorithm $\mathcal{K}$ achieves $\epsilon$-DP if  $$ \frac{P(\mathcal{K}(\bm d) \in R)}{P(\mathcal{K}(\bm d') \in R)} \leq \exp(\epsilon)$$
for all ($\bm d, \bm d'$) and all $R \subseteq \mbox{image}(\mathcal{K})$. This is the typical representation of the bound, which is agnostic to whether $\bm d$ or $\bm d'$ is in the denominator. The parameter $\epsilon > 0$ is chosen by the researcher and can be thought of as a ``privacy budget.'' The smaller the privacy budget $\epsilon$, the less risk of disclosure.

Algorithms that achieve $\epsilon$-DP can be impractical as the outputs may not resemble the restricted statistics closely enough to be useful. Therefore, multiple relaxations have been developed including  $(\epsilon,\delta)$-DP \citep{dwork_differential_2010} which guarantees that:

$$ P(\mathcal{K}(\bm d) \in R) \leq P(\mathcal{K}(\bm d') \in R) \exp(\epsilon) + \delta$$ In other words, $\mathcal{K}$ achieves $\epsilon$-DP with probability $1-\delta$. We will focus on $(\epsilon,\delta)$-DP in this paper. A common algorithm which achieves $(\epsilon,\delta)$-DP is the Gaussian mechanism \citep{dwork_algorithmic_2013}, which adds random Gaussian noise to a statistic (called a query in the literature) calculated from the restricted data. Let $f(\bm d) \in \mathbb{R}^k, k\geq 1$ be a statistic, then $\mathcal{K}(\bm d,\gamma) = f(\bm d) + N_k(0,\gamma^2 \bm I)$. $\mathcal{K}(\bm d,\gamma)$ achieves $(\epsilon,\delta)$-DP when 

$$\gamma =  \Delta f\sqrt{2\log(1.25/\delta)}/\epsilon$$

$$\Delta f =  \max\limits_{\bm d,\bm d'}||f(\bm d') - f(\bm d)||_2.$$
The researcher chooses $\epsilon$ and $\delta$, which together make up the privacy budget. $\Delta f$ is called the global sensitivity of the statistic or query, which is a measure of how much $f(\cdot)$ could possibly change between \textit{any} ($\bm d ,\bm d'$). To calculate the global sensitivity, for many typical statistics, like the empirical mean or empirical variance, one must know or assume bounds on the data $\bm d$. See \cite{bowen_philosophy_2021, barrientos_feasibility_2023} for discussions of calculating global sensitivities.

DP algorithms maintain a couple of benefits. First, the definition makes no assumptions about the information that an attacker has. DP algorithms are also robust to post processing, so any transformation of a differentially private output is still differentially private. DP algorithms also have useful composition properties  \cite{barrientos_feasibility_2023}.  When multiple statistics are calculated from the same data, the amount of privacy budget used for each statistic is simply added  to calculate the total privacy budget used, under Sequential Composition \citep{dwork_algorithmic_2013}. The composition of a $(\epsilon_1, \delta_1)$-DP algorithm and a $(\epsilon_2, \delta_2)$-DP algorithm applied to the same data is a $(\epsilon_1+\epsilon_2, \delta_1 + \delta_2)$-DP algorithm. Therefore, to maintain $(\epsilon, \delta)$-DP among $q$ different statistics $f(\bm d)$ from the same data, a $(\epsilon/q, \delta/q)$-DP algorithm is applied to each statistic. If a $(\epsilon, \delta)$-DP algorithm is repeatedly applied to \textit{disjoint} sets of a dataset (for example, histogram bins), then under Parallel Composition, the result is a $(\epsilon, \delta)$-DP algorithm (so the budget does not need to be split) \citep{mcsherry_privacy_2009}.

There is a clear trade-off between the privacy budget $(\epsilon, \delta)$ and the magnitude of the noise added to the statistic ($\gamma$). The magnitude of the noise added to statistics additionally depends on the sensitivity of statistics to changing one observation in the confidential data (therefore, the largest outlier) and increases with the number of statistics calculated from the data. Therefore, the magnitude of the noise added can be so large that the transformed data is no longer useful, if the privacy budget is small. 

Organizations have started adopting differential privacy for disclosure limitation in recent years \cite{wood_differential_2018}. Notably, the U.S. Census Bureau implemented a differentially private algorithm for releases of the 2020 Census redistricting data with $\epsilon = 17.14, \delta = 10^{-10}$ \citep{jarmin_disclosure_nodate}. There are no guidelines for choosing the privacy budget $\epsilon$ and $\delta$, which is ultimately a policy choice \cite{bowen_philosophy_2021}. In general, organizations currently use large privacy budgets because they make many queries from the same dataset. For example, in 2020 Google had a monthly privacy budget of $\epsilon \approx 80$ for Mobility Reports \cite{rogers_linkedins_2020}.


\subsection{Disclosure Limiting Transformations}\label{sec:priv-transforms}

We will compare two major approaches to releasing confidential data while limiting disclosure risk -- synthetic data (\ref{sec:priv-syn}) and a differentially private Gram matrix (\ref{sec:priv-gram}) -- as described in the following. 

\subsubsection{Synthetic Data}\label{sec:priv-syn}

Synthetic data, introduced by \cite{rubin_discussion_1993}, is a synthetic version of a dataset when raw microdata cannot be released. \cite{rubin_discussion_1993} and \cite{little_statistical_1993} viewed synthetic data as a missing data problem, where the sensitive information was missing and could be imputed with multiple imputation. In general, synthetic data replaces sensitive information in the original data with values generated from statistical summaries of the original data \cite{snoke_general_2018, raghunathan_synthetic_2021}. The big idea, typically, is to generate synthetic data by sampling from the empirical joint and marginal distributions of the columns in the confidential data. For example, a popular parametric method for data synthesis generates variables sequentially, generating the next variable with predictions from classification and regression trees (CART)  fit on the already synthesized variables \citep{nowok_synthpop_2016}. We use this synthesis method in the simulations in this paper. 

Synthetic data is appealing as a disclosure limiting transformation of confidential data because it can technically be analyzed with the same methods as the original data. However, it has two major drawbacks. First, even though no observations from the confidential data are released in such synthetic data, there is still a risk of information disclosure. Recent work has discussed the risk of leaking information with typical synthetic data generation \cite{bellovin_privacy_2018}. More recently, methods have been developed to generate differentially private synthetic data \cite{wilchek_synthetic_2021,boedihardjo_private_2022}, in which case the disclosure risk is clear, via the choice of privacy budget.  The second drawback of releasing synthetic data is that making valid inferences with synthetic data requires clear communication from the data steward to the public of how the data can be analyzed. 

\subsubsection{Noise Infused Gram Matrix}\label{sec:priv-gram}

The major benefit of synthetic data is that it is at the same observation level as the confidential data that cannot be released. However, the estimators discussed in Section~\ref{sec:notation} do not require the individual level auxiliary data. Estimating the PATE with $\tcw(\cdot)$, only requires the first or second empirical moments of the auxiliary data. Estimating the SATE with $\tregy$, only requires predictions from a model fit of the covariates on the outcome of interest with the auxiliary data. Therefore, we consider statistical summaries of the auxiliary data to limit disclosure risk.

We define the data matrix $\bm D_{m \times p +2}$ with observations $\bm D_i = (1, Y_i, \bXi)$. Let $\bm Y$ denote the $m \times 1$ vector of observed outcomes, $\bX$ denote the $m \times p$ matrix of covariates, and $\bm 1$ denote a $m \times 1$ vector of 1's.  Then, the gram matrix of the data matrix $\g = \bm D ' \bm D/m$ includes $\bX ' \bX/m$, $\bX ' \bm Y/m$, $ \hat{\mu}_Y =\bm 1 ' \bm Y/m$, and $\hat{\bm \mu}_x=  \bm 1 ' \bX/m$.  The coefficients $\bm \beta$ for an OLS model are estimated as $\hat{\bm \beta} = (\bX'\bX)^{-1}\bX ' \bm Y$, so $\g$ is sufficient to estimate $\hat{\bm \beta}$ for the auxiliary study to leverage in $\tregy$.  $\g$ includes the first and second empirical moments of $\bm D$, to leverage in $\tcw(\cdot)$.

Therefore, the gram matrix of the auxiliary data $\g$ seems like a sensible place to start for a disclosure limiting transformation of the data. In fact, $\g = \bm D ' \bm D/m$ is a special version of matrix masking, a SDC technique \cite{duncan_enhancing_1991}. However, this type of matrix masking does not meet current standards for disclosure limitation, so we propose releasing a noisy version of the gram matrix of the auxiliary data. Specifically, we consider a $(\epsilon,\delta)$-DP algorithm for releasing a noise infused gram matrix.

There is a body of literature that considers perturbed sufficient statistics for linear regression (OLS), penalized regression (such as Ridge and LASSO), and Principle Component Analysis (PCA), which are all contained in $\bm G$. See \cite{barrientos_feasibility_2023} for a recent review.
\cite{kamath_privately_2019}, \cite{balle_improving_2018},  and \cite{wang_revisiting_2018}, building off of the work in \cite{dwork_analyze_2014},  propose methods for adding Gaussian noise to a covariance matrix to achieve $(\epsilon, \delta)$-DP. \cite{chanyaswad_mvg_2018} considers matrix-valued DP releases in general and illustrates a method adding multivariate-Gaussian noise with the Gram matrix.
Another line of work proposes adding Laplace noise to the covariance matrix to achieve $\epsilon$-DP \citep{ferrando_parametric_2021, jiang_wishart_2016, foulds_theory_2016, vu_differential_2009, alabi_differentially_2020}. \cite{sheffet_old_2019} proposes a mechanism for adding Wishart noise to achieve $(\epsilon, \delta)$-DP. Most of these approaches do not address the issue of variables in real datasets having largely different scales.  Therefore, we take a similar approach to \cite{dwork_analyze_2014}, but adapt it to account for such differences in scale.

$\g$ is a symmetric matrix of statistics calculated from the data $\bm D$. Therefore, we can think of the upper triangle of $\g$ as a vector of statistics with length $k = \frac{1}{2}(p^2 + 5p +2)$ and could apply the Gaussian mechanism as described in Section~\ref{sec:priv-dp} to this vector with some privacy budget $(\epsilon, \delta)$ as in \cite{dwork_analyze_2014}.  However, the magnitude of the Gaussian noise added to each element of the vector depends on the largest difference of the upper triangle of $\g$ given one observation is removed. The columns of $\bm D$ might be on vastly different scales and with different variances. This is therefore not the best approach to attain $(\epsilon, \delta)$-DP while adding as little noise as possible. 

\begin{figure}[ht]
    \centering
    \includegraphics[width = .3\textwidth]{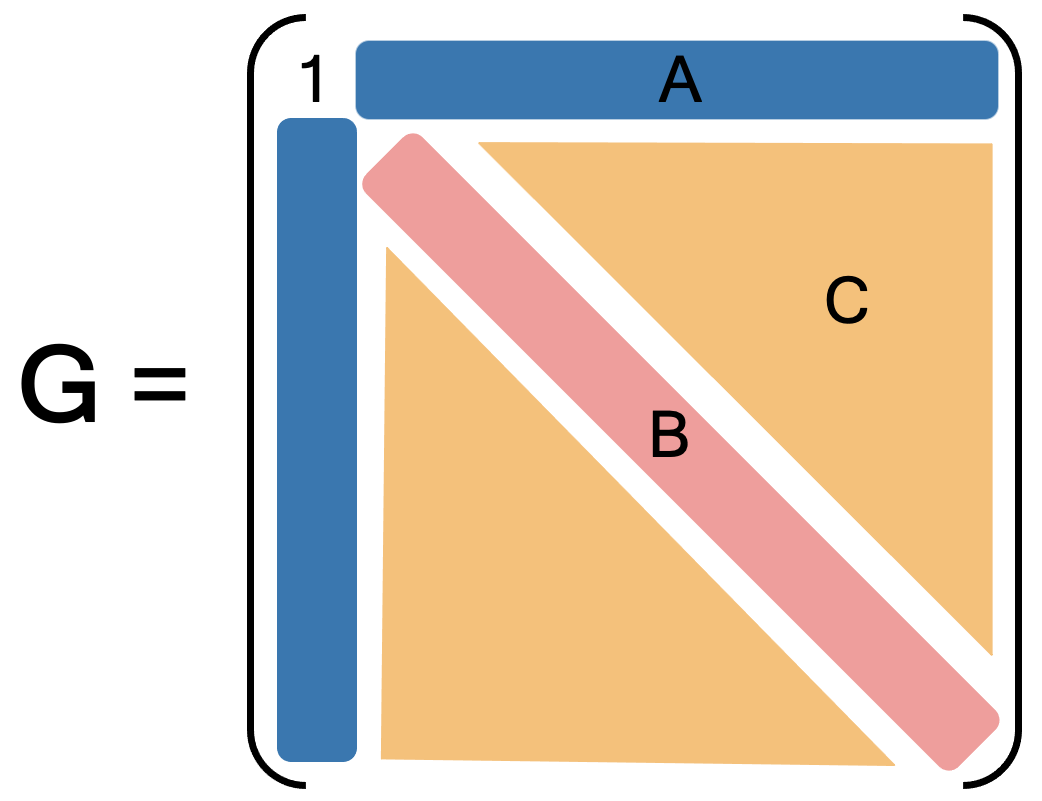}
    \caption{Illustration of how the gram matrix $\g$ can be partitioned into separate parts, which correspond to the column means (A), empirical second moments (B) and empirical moments of the columns cross multiplied (C).}
    \label{fig:g}
\end{figure}

Instead, we divide the upper triangle of $\g$, including the diagonal, into different elements, as illustrated in Figure~\ref{fig:g}, and divide the privacy budget among those elements.  First, we assume that the sample size $m$ is not confidential. Let $\bm 1$ denote a $m\times1$ vector of 1's. Define the vector of column means, $\hat{\bm \mu}= \bm 1'\bm D/m$, the empirical moments of the columns cross-multiplied correlation matrix $\hat{\bm C}_{p+1 \times p+1} = (\bm Y,\bm X)'(\bm Y,\bm X)$, and the vector of empirical second moments $\hat{\bm \mu}_2= \mbox{diag}(\hat{\bm C})$. Then, $\bm G$ can be reconstructed as illustrated in  Figure~\ref{fig:g} with A = $\hat{\bm \mu}$, B = $\hat{\bm \mu}_2$ and C is the upper triangle of $\hat{\bm C}$ (excluding the diagonal).  

We construct $(\epsilon, \delta)-$DP $\g$ as follows. First, we divide the privacy budget between $\hat{\bm \mu}$, $\hat{\bm \mu}_2$, and the upper triangle of $\hat{\bm C}$ proportionally to the number of elements in each. Therefore, $\hat{\bm \mu}$ and $\hat{\bm \mu}_2$ are allocated $\frac{2(p+1)}{(p+2)(p+3)-2}$ and the upper triangle of $\hat{\bm C}$ is allocated $\frac{p(p+1)}{(p+2)(p+3)-2}$ of the privacy budget. We apply the Gaussian mechanism to each element of $\hat{\bm \mu}$, $\hat{\bm \mu}_2$, and the upper triangle of $\hat{\bm C}$ separately. Therefore, we split the privacy budget for the upper triangle of $\hat{\bm C}$ across the $p(p+1)/2$ elements to construct a $\big(\frac{p(p+1)}{(p+2)(p+3)-2} \epsilon, \frac{p(p+1)}{(p+2)(p+3)-2} \delta\big)$-DP release of the upper triangle of $\hat{\bm C}$. By parallel composition, we do not need to further split the privacy budget to construct $\big(\frac{2(p+1)}{(p+2)(p+3)-2} \epsilon, \frac{2(p+1)}{(p+2)(p+3)-2} \delta\big)$-DP releases of $\hat{\bm \mu}$ and $\hat{\bm \mu}_2$. Finally, we reconstruct a DP gram matrix, $\gstar$, from the DP releases of  $\hat{\bm \mu}$ and $\hat{\bm C}$. Due to sequential composition, the resulting noisy gram matrix $\gstar$ is $(\epsilon, \delta)-$DP. 

Dividing the gram matrix into these elements allows adding a smaller magnitude of noise while still achieving differential privacy. First, separating the matrix into the different elements accounts for the different sensitivities of first and second moments, and second, applying the Gaussian mechanism element-wise within each element accounts for scale differences between columns. See Appendix~\ref{appx:dpg-matrix} for details of the sensitivity calculations that are used in the algorithm.

There is no guarantee that the resulting DP matrix will be positive definite -- an important property of gram matrices. Therefore, we post-process $\gstar$ to ensure that it is positive-definite in a similar manner to \cite{barrientos_feasibility_2023}. Namely, we set any negative eigenvalues to zero, add the median positive eigenvalue to all of the eigenvalues, and then reconstruct the matrix with the original eigenvectors and the transformed eigenvalues.

\section{Simulation Studies}\label{sec:sims}


We conduct two simulation studies to compare the utility of different auxiliary data releases in estimators of the PATE and SATE respectively. As baselines, we use the difference-in-means estimator $\hat{\tau}^{DM}$ and regression estimator $\tregrct$, which only rely on RCT data. The difference-in-means estimator is the difference in the mean outcomes for the subjects assigned treatment and those assigned control. These are compared to the calibration weighted estimator and super-covariate data integration approach using regression, as discussed in Sections~\ref{sec:priv-syn} and~\ref{sec:priv-gram}.


\subsection{Generalizing to Populations of Interest}\label{sec:sim-gen}

We consider a setting where there is an RCT and a related observational study which were both sampled from a population of interest. The aim is to estimate the PATE for this population of interest. The auxiliary study is representative of this population. However, the RCT sample is not representative of the population of interest due to selection bias.  We consider a modified version of the simulations in \cite{colnet_causal_2021} and \cite{lee_improving_2021}, which assumes a heterogeneous treatment effect. 

\subsubsection{Data Generation}

We emulate a hypothetical randomized experiment and auxiliary study assuming that the SATE in the RCT does not equal the desired PATE due to selection bias. We generate the $1 \times p$ covariate vector $\bXi$ from i.i.d $N(\bm 1,\bm I_p)$ distributions. We additionally generate a covariate $X^S_i \sim N(1,1)$, which impacts both selection into the RCT and subject $i$'s treatment effect. Let $S_i$ be an indicator of whether subject $i$ is selected into the RCT and $\pi_i^S = P(S_i = 1)$. We model the probability of selection into the RCT with a logistic regression model: $$\mbox{logit}\{\pi_i^S\} = -2 + \bm \beta_S\bXi + .5X^S_i.$$ $\bm \beta_S$ is generated for each value of $p$, then fixed, with 50\% of the elements of $\bm \beta_S$ set to be $-1/(.5 p)$ and the other are 0. Then $S_i$ is generated from a Bernoilli distribution with probability $\pi_i^S$. If $S_i = 1$, subject $i$ is included in the RCT sample. We generate $\bXi$, $X^S_i$, and $S_i$ $1,300$ times so that approximately 100 subjects are selected into the RCT ($n \approx 100$). The auxiliary study sample ($m = 10,000$) is then generated directly from the population, with $\bXi \sim N(\bm 1, \bm I_p)$ and $X^S_i \sim N(1,1)$. The control potential outcomes are generated  $Y_i^c = .5 + \beta\bXi + \varepsilon_i$, $\varepsilon_i \sim N(0,\sqrt{.3})$. $\beta$ is fixed for each value of $p$, with 60\% of the elements randomly selected to be $\sqrt{.7}/(.6p)$ and the others are 0. Therefore, some covariates contribute to both the $\pi_i^S$ and the outcome, one, or neither and the covariates explain 70\% of the variance in $Y_i^c$. We let $Y_i^t = Y_i^c + .5X^S_i$. Since $\mathbb{E}[X^S] = 1,$ $\tau_{PATE} = .5$. Based on the selection model, higher values of $X^S_i$ are favored for selection into the RCT so the SATE will be larger than the PATE. 

\subsubsection{Simulation Procedure}\label{sec:sim-gen-gen}

For $p=10$ and $p=20$, we repeat the following procedure 1,000 times. First, we generate an observational and experimental dataset as described in the previous section. We then calculate each disclosure limiting transformation of the auxiliary data as described in Section~\ref{sec:priv-transforms}, and the mean vector for the $p+1$ covariates $\{\bm X, \bm X^S\}$ from each transformation. Denote the mean vector $\hat{\mu}(\cdot)$, so $\hat{\mu}(\g)$ is the (true) mean vector calculated from $\g$. We implement the differentially private algorithm (\ref{sec:priv-gram}) to generate $\gstar$ with a range of $\epsilon$ between 1 and 30 and $\delta = 10^{-5}$. We additionally generate a synthetic dataset ($\syn$) using the \texttt{synthpop} package in \texttt{R} \cite{nowok_synthpop_2016}, which implements sequential synthesis.

\begin{figure}[]
    \centering
    \input{figures/gen-mse-comb-paper-final}
    \caption{MSE of estimators of the PATE calculated across 1,000 iterations  for different numbers of covariates ($p$). Error bars represent two simulation standard errors. The MSE is decomposed into the squared bias and the variance of the estimator. The first two estimators do not use data integration: $\tdm$ is the difference-in-means estimator and $\tregrct$ is the the regression estimator with only RCT covariates. We do not show results for $\tacwt$ with $\gstar$, $\epsilon = 1$ and $p = 20$ due to computational in-feasibility.  }
    \label{fig:gen-mse}
\end{figure}

We then generate the treatment assignment $T_i \sim \mbox{Bern}(.5)$ for the RCT sample, calculate the observed outcomes $Y_i = T_iY_i^t + (1-T_i)Y_i^c$, and calculate the estimators. To compare the difference-in-means and regression estimators to a calibrated weighted estimator with more comparable variance, we calculate the augmented CW estimator $\tacwt(\cdot)$ as described in \cite{lee_improving_2021}. The weights are estimated to calibrate the empirical mean of the RCT to the empirical mean of the auxiliary study $\hat{\mu}(\cdot)$; i.e. in the optimization problem (Equation~\ref{eq:calib}), $\frac{1}{m} \sum_{i \in \mathcal{O}}g(\bm{X}_i) = \hat{\mu}(\cdot)$. The weights are estimated by solving the optimization problem with Lagrange multipliers and Newton's equation \citep[see][for more details]{lee_improving_2021}. We adapt the \texttt{genRCT} package \cite{yang_genrct_2021} code to calculate $\tacwt(\cdot)$ in \texttt{R}. We additionally estimate a 95\% confidence interval for each estimator. We use the standard variance estimators for the difference-in-means and regression estimators and the bootstrap variance estimator suggested by \cite{lee_improving_2021} with 100 bootstraps for the $\tacwt(\cdot)$ estimator.

\subsubsection{Utility Metrics}

We evaluate the utility of each private data release for estimating the PATE with two metrics. First, we look at the Mean Squared Error (MSE) of the estimator. Second, we consider the coverage probability for a 95\% confidence interval. We additionally consider utility metrics for estimates of the column means themselves in the auxiliary data. For this, we compare the square root of the MSE (RMSE) of the column means for each disclosure limiting transformation to the RMSE of the confidential auxiliary data ($\g$). 

\subsubsection{Results}

Figure~\ref{fig:gen-mse} shows the empirical MSE for each estimator, estimated across the 1,000 simulations. We note that we do not show results for the DP release $\gstar$ with $\epsilon = 1$ and $p=20$  because the algorithm could not find a solution to the calibration weights in a large portion of the simulations. Blue represents the variance component of MSE and orange represents the squared bias component of the MSE. The difference-in-means and regression estimators, which only rely on data from the RCT have large estimated MSEs, which are primarily due to large squared biases, as compared to little or no bias of $\tacwt(\cdot)$. Due to the shift in the distribution of $X^S$ in the RCT sample, the average SATE across simulations is .86, which larger than the true PATE of .5. The difference-in-means and regression estimators are unbiased for the SATE, so they are upwardly biased for the PATE in this case. $\tacwt(\cdot)$ performs similarly across different disclosure limiting transformations of the auxiliary data. The CW estimator using the DP Gram matrix releases, even with a small privacy budget (corresponding to high levels of privacy) performs similarly to the original auxiliary data when $p=10$. The results are similar when $p = 20$, although the variances of all estimators increases. 

\begin{table}[]
    \centering
\centering
\resizebox{\ifdim\width>\linewidth\linewidth\else\width\fi}{!}{
\begin{tabular}{lccc}
\toprule
\multicolumn{2}{c}{\textbf{ }} & \multicolumn{2}{c}{\textbf{Coverage}} \\
\cmidrule(l{3pt}r{3pt}){3-4}
\multicolumn{1}{c}{\textbf{Estimate of the PATE}} & \multicolumn{1}{c}{\textbf{$\epsilon$}} & \multicolumn{1}{c}{\textbf{p = 10}} & \multicolumn{1}{c}{\textbf{p = 20}}\\
\midrule
\addlinespace[0.3em]
\multicolumn{4}{l}{\textbf{RCT data only}}\\
\hspace{1em}$\tdm$ & - & 0.51 (0.010) & 0.51 (0.013)\\
\hspace{1em}$\tregrct$ & - & 0.08 (0.009) & 0.13 (0.010)\\
\addlinespace[0.3em]
\multicolumn{4}{l}{\textbf{Includes auxiliary data}}\\
\hspace{1em}$\tacwt(\g)$ & - & 0.95 (0.007) & 0.96 (0.005)\\
\hspace{1em}$\tacwt(\gstar)$ & 1 & 0.94 (0.007) & -\\
\hspace{1em}$\tacwt(\gstar)$ & 3 & 0.95 (0.007) & 0.92 (0.012)\\
\hspace{1em}$\tacwt(\gstar)$ & 6 & 0.95 (0.006) & 0.95 (0.008)\\
\hspace{1em}$\tacwt(\gstar)$ & 15 & 0.96 (0.005) & 0.95 (0.007)\\
\hspace{1em}$\tacwt(\gstar)$ & 30 & 0.95 (0.005) & 0.95 (0.005)\\
\hspace{1em}$\tacwt(\syn)$ & - & 0.95 (0.006) & 0.95 (0.006)\\
\bottomrule
\end{tabular}}

 \caption{Estimated coverage probability for 95\% confidence intervals for estimators of the PATE for different numbers of covariates ($p$). Simulation standard errors in parentheses.}
    \label{tab:gen-coverage}
\end{table}

Table~\ref{tab:gen-coverage} shows the empirical coverage probability of the true PATE (.5) for 95\% confidence intervals, across the 1,000 simulations. The regression estimator has very poor coverage for the PATE, as does the difference-in-means estimator, which only has higher coverage because the variance is larger. The coverage of bootstrap 95\% confidence intervals for the calibration weighted estimator is close to .95 cross the different disclosure limiting transformations of the auxiliary data, with the exception of $\gstar$ (DP Gram matrix) when $\epsilon = 1$ and when $\epsilon = 3$ and $p = 20$.

Table~\ref{tab:rmse-gen} shows the RMSE of the column means for the different releases of the confidential auxiliary data. This gives some insight into the utility of releasing $\gstar$ with different privacy budgets $\epsilon$ or synthetic data for uses of the column means other than the data integration estimator considered here. The RMSE of the column means for synthetic data is larger than the confidential data but only changes a small amount when the number of covariates increases. We find that when $p=10$, for $\epsilon \geq 6$, the RMSE of the column means is essential the same as the original confidential data. However, there is a large jump in the RMSE when $\epsilon = 1$ and for small privacy budgets when there are more covariates ($p = 20$). 

To summarize, when the PATE is the estimand of interest, leveraging auxiliary data in the analysis of an RCT with $\tacwt(\cdot)$ can greatly reduce the MSE of the treatment effect estimate, and this continues to be true when using auxiliary data that has undergone disclosure limiting transformations.

\begin{table}[]
    \centering
    \centering
\resizebox{\ifdim\width>\linewidth\linewidth\else\width\fi}{!}{
\begin{tabular}{lccc}
\toprule
\multicolumn{2}{c}{\textbf{ }} & \multicolumn{2}{c}{\textbf{RMSE of Column Means} ($\times 10^{-2}$)} \\
\cmidrule(l{3pt}r{3pt}){3-4}
\multicolumn{1}{c}{\textbf{Data Release}} & \multicolumn{1}{c}{\textbf{$\epsilon$}} & \multicolumn{1}{c}{\textbf{p = 10}} & \multicolumn{1}{c}{\textbf{p = 20}}\\
\midrule
$\g$ (No privacy) & - & 0.98 (0.007) & 0.99 (0.005)\\
$\syn$  & - & 1.44 (0.009) & 1.51 (0.008)\\
$\gstar$ (DP) & 1 & 8.58 (0.089) & -\\
$\gstar$ (DP) & 3 & 2.02 (0.026) & 8.16 (0.059)\\
$\gstar$ (DP) & 6 & 1.09 (0.009) & 3.80 (0.030)\\
$\gstar$ (DP) & 15 & 0.97 (0.007) & 1.29 (0.009)\\
$\gstar$ (DP) & 30 & 0.97 (0.007) & 1.00 (0.005)\\
\bottomrule
\end{tabular}}

    \caption{RMSE of column means for releases of confidential auxiliary (observational) data, varying the number of covariates $p$. Simulation standard errors in parentheses.}
    \label{tab:rmse-gen}
\end{table}

\subsection{Improving Experimental Precision}\label{sec:sim-var}

In the second simulation study, the goal is to estimate the average treatment effect for an RCT sample. We assume that there is an RCT evaluating a specific treatment, and a related observational study which includes the outcome of interest. We do not necessarily observe the treatment in the auxiliary observational study. However, given that observational studies typically have large sample sizes, we can expect to get good estimates of model parameters for a model of the outcome on observed covariates, resulting in good predictions of the outcome. Further, if there are a large number of covariates, an observational study will likely better estimate model parameters for all covariates than a small RCT. As discussed in Section~\ref{sec:prec_est}, predictions of the outcome based on a model trained on the auxiliary data can therefore be a very powerful covariate to include in RCT covariate adjustment. For simplicity, we consider a setting where the RCT and observational samples arise from the same, linear, data generating model (there is no covariate shift) and where the regression models match this data generating model. There are other covariate adjustment methods incorporating auxiliary data that account for covariate shift (see Appendix~\ref{appx:reloop}).

\subsubsection{Data Generation}
We emulate a hypothetical RCT with $n=100$ and an auxiliary (observational) data set with $m = 10,000$. For both, we generate $\bXi$ from i.i.d $N(\bm 0, \bm I_p)$ distributions. Then, $Y_i^c = .5 + \beta' \bXi + \varepsilon_i$, where $\varepsilon_i \sim N(0,.3)$. To make some covariates contribute to the outcome, while others are noise, .6 of the elements of $\bm \beta_{1 \times p}$ are $\sqrt{.7/(.6p)}$, while the rest are set to 0. By this data generating process, $\mathbb{V}[Y_i^c]=1$, and the proportion of variance explained by the covariates is .7. We let $\tau_{SATE}= .5$, so $Y_i^t = Y_i^c + .5$ under the constant treatment effect assumption. In the auxiliary data, we assume that there is no treatment, so $Y_i = Y_i^c$ for $i\in \mathcal{O}$.

\subsubsection{Simulation Procedure}

For each value of $p$, we implement the below 100 times (100 data generations).

First, we generate an RCT and auxiliary data set as described above. We then calculate each disclosure limiting transformation of the auxiliary data as described in the previous section (\ref{sec:sim-gen-gen}). With each transformation of the auxiliary data, we calculate the OLS estimated coefficients for a model of the outcome on the covariates (auxiliary model). With each of these coefficient vectors, we predict the outcome for subjects in the RCT ($i\in\mathcal{R}$) (auxiliary predictions). Denote this prediction $\yhati$, so $\hat{Y}_i(\g)$ is the prediction of the outcome for subject $i$ based on the coefficients calculated from $\g$. Let $\yhato(\g)$ denote the $n \times 1$ length vector of these predictions for the RCT subjects. The auxiliary data is set aside. 

Since we are interested in the average treatment effect for an RCT sample, we treat the outcomes and covariates as fixed. We then generate 1,000 treatment assignment vectors $T_i \sim \mbox{Bern}(.5),i\in\mathcal{R}$. For each treatment assignment vector we generate the observed outcomes and calculate each estimator: $\tdm$, $\tregrctf$, and $\tregyaux$ for each disclosure limiting transformation. Thus, we get an estimate of the distribution of the comparison estimators, based on a simulated distribution of the treatment assignment. Because $n = 100$, the regression estimator fit only on the RCT covariates, $\tregrctf$, runs into dimensionality issues as $p$ grows. Therefore, we allow a maximum of 20 RCT covariates to be included in the regression model. With $p=50$, we choose 20 covariates that have a non-zero coefficient in the data generating model, emulating analyses of the RCT using variable selection and/or expert knowledge that would select predictive covariates.

\subsubsection{Utility Metrics}

Because we use unbiased estimators for the SATE, we evaluate the utility of the data integration approach, with different auxiliary data releases, using the variance of the given point estimator. We calculate the empirical variance of the estimates across the 1,000 treatment assignment vectors for each data generation for this comparison. To more easily compare the variances, we consider the \textit{relative efficiency }of each point estimator as compared to  $\tregrct$, which only relies on RCT data. The relative efficiency is calculated as the ratio of the the (simulated) variance of $\tregrct$ and the variance of each estimator. The relative efficiency can also be interpreted as a multiplicative effect on the required sample size for a given estimator to achieve the same efficiency as $\tregrct$. Relative efficiencies greater than 1 indicate that the corresponding estimator is more precise than $\tregrct$, while relative efficiencies less than 1 indicate the opposite. For the main results, we average the variances and relative efficiencies of each estimator across the 100 data generations and calculate a standard error as the standard deviation of the 100 metrics divided by $\sqrt{100}$.  

We additionally consider utility for statistical summaries of the auxiliary data releases, which are the intermediate step to the data integration method. First, we use look at RMSE of the coefficient vector $\hat{\bm \beta}$ calculated from an auxiliary data release. This is a typical utility metric used for OLS in the DP literature \cite{barrientos_feasibility_2023}. Finally, we consider the Frobenius and spectral norms of the difference between the confidential Gram matrix $\g$ and the Gram matrix resulting from each transformation.

\begin{figure}
    \centering
    \input{figures/re-regression-paper-final}
    \caption{Simulated relative efficiency of $\treg(\cdot)$ estimator with different covariates compared with the $\tregrct$ estimator (using RCT covariates only). The relative efficiency, or sample size multiplier, is calculated as $\tregrct/\treg(\cdot)$, so values larger than 1 mean that the estimator is more efficient. $\yhato(\cdot)$ is the vector of predictions for the RCT sample from a model fit on the auxiliary data using a certain release of the auxiliary data. $\epsilon$ is the privacy budget for DP transformations. Error bars represent two simulation standard errors.}
    \label{fig:var-sims-reg}
\end{figure}

\subsubsection{Results}

Figure~\ref{fig:var-sims-reg} shows the relative efficiency of each estimator compared to $\tregrct$, for $p = 10, 20, \mbox{and } 50$ covariates. Let's first focus on the performance of the super covariate data integration approach if we had access to the confidential auxiliary data ($\g$). Since the regression model fit with the RCT was correctly specified for $p=10 \mbox{ and } p = 20$, we expected there to be only small gains in precision integrating the observational data. When $p=50$, the regression estimator fit on only the RCT covariates is limited in the number of predictive covariates that can be included in the model, so there are greater gains to using predictions fit on the auxiliary data.  All of the estimators fall somewhere between the difference-in-means estimator ($\emptyset$, orange circle) and the data integration approach with the confidential auxiliary data ($\g$, purple triangle).

Using synthetic data ($\syn$, blue square) in the super-covariate data integration approach performs similarly to using the confidential data itself ($\g$), always outperforming using the RCT covariates alone.  The performance of the DP Gram matrix release ($\gstar$, shades of green triangles) depends on the number of covariates and the privacy budget, $\epsilon$. When there are a small number of covariates $p = 10$, using $\yhato(\gstar)$ as a covariate in the regression estimator performs as well as using the RCT covariates when $\epsilon$ is greater than 6. However, when $p = 20$ a privacy budget of $\epsilon = 15$ or greater is needed for the data integration approach to outperform regression with the RCT covariates alone, using $\yhato(\gstar)$. When $p= 50$ the data integration approach with the DP Gram matrix never outperforms using the RCT covariates alone, and for most of the privacy budgets, it has similar variance to the difference-in-means estimator. The DP transformations lose utility quickly as $p$ increases because the required magnitude of noise for the Gaussian mechanism increases with $p^2$. These results clearly illustrate the privacy-utility trade-off, with the relative efficiency of $\treg(\yhato(\gstar))$ increasing as the privacy budget increases. 

\begin{figure}
    \centering
    \input{figures/auxsize-reg-final}
    \caption{Simulated variance for $\treg(\cdot)$ with $p = 10$ covariates using only RCT data (circles) and incorporating auxiliary information (triangles), increasing the size of the auxiliary data, $m$. $\yhato(\cdot)$ is the vector of predictions for the RCT sample from a model fit on the auxiliary data using a certain release of the auxiliary data. $\epsilon$ is the privacy budget for DP transformations. Error bars represent two simulation standard errors.}
    \label{fig:auxsize}
\end{figure}

We find that the utility of using the super-covariate data integration is lost for the DP transformations of auxliary data, when there are a large number of covaraites. However, the noise necessary to achieve DP additionally depends on the sensitivity of the statistics to removing one observation from the data. Thus, the sensitivity decreases as $m$ increases. Figure~\ref{fig:auxsize} shows the variance of $\treg$ with different covariates, varying the size of the auxiliary study. It illustrates that the utility of $\yhato(\gstar)$ as a covariate stays consistent with the utility of $\yhato(\g)$ as the size of the auxiliary study gets very large.

As noted previously, these results are based on correctly specified regression models and RCT and auxiliary samples that arise from the same data generating process. In practice, neither of these assumptions may be true, and it can be preferred to use design-based estimators. See Appendix~\ref{appx:reloop} for a discussion and simulations which show that these results hold with a design-based estimator that is robust to model mis-specification and covariate shift.

Table~\ref{tab:utility} presents additional utility metrics for the disclosure limiting releases. As with Table~\ref{tab:rmse-gen}, this can give us an idea of the utility of the releases for uses of the Gram matrix aside from the super-covariate data integration approach. First, in terms of the OLS estimated coefficients for a regression model with the auxiliary sample ($\hat{\beta}$), the synthetic data ($\syn$) performs very similarly to the original, confidential, auxiliary data ($\g$). When $p = 10$ and $\epsilon > 6$, the DP Gram matrix also estimates the coefficients with similar RMSE. The RMSE of $\hat{\beta}$ is much larger for the DP gram matrices ($\gstar$) with a large number of covariates and small privacy budgets, than the confidential data ($\g$). There is a similar pattern with the Frobenius and spectral norms of the different Gram matrix releases.

\begin{table}[ht]
\centering
\centering
\resizebox{\ifdim\width>\linewidth\linewidth\else\width\fi}{!}{
\begin{tabular}{lcccc}
\toprule
\multicolumn{2}{c}{ } & \multicolumn{1}{c}{$\hat{\bm \beta}$} & \multicolumn{2}{c}{\textbf{vs. Non-Private Gram Matrix}} \\
\cmidrule(l{3pt}r{3pt}){3-3} \cmidrule(l{3pt}r{3pt}){4-5}
\multicolumn{1}{c}{\textbf{Release}} & \multicolumn{1}{c}{\textbf{$\epsilon$}} & \multicolumn{1}{c}{\textbf{RMSE}} & \multicolumn{1}{c}{\textbf{Frobenius Norm}} & \multicolumn{1}{c}{\textbf{Spectral Norm}}\\
\midrule
\addlinespace[0.3em]
\multicolumn{5}{l}{\textbf{P = 10}}\\
\midrule
\hspace{1em}$\g$ (No privacy) & - & 0.01 & 0.00 &  0.00\\
\hspace{1em}$\syn$ & - & 0.02 & 0.17 & 0.10\\
\hspace{1em}$\gstar$ (DP) & 1 & 0.24 & 5.69 & 3.05\\
\hspace{1em}$\gstar$ (DP) & 3 & 0.16 & 2.18 & 1.02\\
\hspace{1em}$\gstar$ (DP) & 6 & 0.08 & 0.80 & 0.39\\
\hspace{1em}$\gstar$ (DP) & 15 & 0.03 & 0.25 & 0.14\\
\hspace{1em}$\gstar$ (DP) & 30 & 0.02 & 0.14 & 0.08\\
\hspace{1em}$\gstar$ (DP) & 50 & 0.01 & 0.11 & 0.06\\
\addlinespace[0.3em]
\midrule
\multicolumn{5}{l}{\textbf{P = 20}}\\
\midrule
\hspace{1em}$\g$ (No privacy) & - & 0.01 & 0.00 &  0.00\\
\hspace{1em}$\syn$ & - & 0.03 & 0.34 & 0.19\\
\hspace{1em}$\gstar$ (DP) & 1 & 0.22 & 28.20 & 12.03\\
\hspace{1em}$\gstar$ (DP) & 3 & 0.18 & 10.30 & 4.36\\
\hspace{1em}$\gstar$ (DP) & 6 & 0.15 & 6.30 & 2.48\\
\hspace{1em}$\gstar$ (DP) & 15 & 0.11 & 2.33 & 0.84\\
\hspace{1em}$\gstar$ (DP) & 30 & 0.05 & 0.72 & 0.30\\
\hspace{1em}$\gstar$ (DP) & 50 & 0.03 & 0.44 & 0.19\\
\addlinespace[0.3em]
\midrule
\multicolumn{5}{l}{\textbf{P = 50}}\\
\midrule
\hspace{1em}$\g$ (No privacy) & - & 0.01 & 0.00 & 0.00\\
\hspace{1em}$\syn$ & - & 0.03 & 0.76 & 0.32\\
\hspace{1em}$\gstar$ (DP) & 1 & 0.15 & 373.32 & 105.51\\
\hspace{1em}$\gstar$ (DP) & 3 & 0.14 & 125.09 & 35.37\\
\hspace{1em}$\gstar$ (DP) & 6 & 0.14 & 63.09 & 17.93\\
\hspace{1em}$\gstar$ (DP) & 15 & 0.13 & 26.15 & 7.42\\
\hspace{1em}$\gstar$ (DP) & 30 & 0.12 & 14.18 & 3.97\\
\hspace{1em}$\gstar$ (DP) & 50 & 0.10 & 9.79 & 2.60\\
\bottomrule
\end{tabular}}

\caption{Additional utility metrics for disclosure limiting transformations of confidential auxiliary data. $\hat{\bm\beta}$ is the coefficient vector from an OLS model of the outcome on coefficients. The matrix norms are the norms of the difference between the non-private Gram matrix $\g$ and the corresponding release. We calculate a Gram matrix from $\syn$ to compute the norms.}
\label{tab:utility}
\end{table}

In summary, the utility of the super-covariate data integration approach is more impacted by the type of disclosure limiting transformation applied to the auxiliary data than the CW estimator. This makes sense because this data integration approach requires the full Gram matrix, with $\frac{1}{2}(p^2 + 5p + 2)$ elements with random noise added to them, versus only the mean vector of $p$ elements. If a large privacy budget $\epsilon$ is acceptable, then releasing a DP Gram matrix for the purpose of data integration to improve efficiency with this approach could still be effective. 

\section{Discussion}

In this paper, we presented disclosure limiting transformations of observational data that can be combined with experimental data to estimate the PATE and SATE. These disclosure limiting transformations included a differentially private Gram matrix and synthetic data. We found that leveraging these transformed versions of observational data to estimate the PATE greatly improved the MSE (and eliminated bias) as compared to methods using only the RCT data. We also found that transformed versions of auxiliary data could improve precision when estimating the SATE, beyond the precision achieved through covariate adjustment with RCT covariates alone, if there are few covariates or a large privacy budget is used for the DP Gram matrix release.

There is not a broad discussion of using the Gram matrix of a data matrix as a disclosure limiting transformation in the literature, to the best of our knowledge. The Gram matrix is a reasonable place to start for disclosure limiting data releases since it provides useful summary statistics of the data and some protection against disclosure, which can be augmented with additional random noise. The Gram matrix also supports some flexible model fitting -- any outcome and covariates can be chosen out of the columns in the data matrix. As has been discussed in the DP literature, a DP covariance matrix can be used for PCA, ridge, and LASSO regularized regression in addition to standard OLS.

In addition to disclosure risk and utility, there are practical considerations that could be taken into account when choosing a disclosure limiting technique. An important consideration is that the additional uncertainty introduced by privacy-transformations may be more challenging to communicate to data users for some techniques than for others. 
In the case of the DP Gram matrix release, $\gstar$, the additional uncertainty is rather clear -- we add noise with a certain distribution to the Gram matrix, the variance of which is parameterized with a couple of parameters. On the other hand, estimating valid standard errors associated with synthetic data requires multiple replicates of the synthetic data and survey sampling methods.

We focus on a small set of disclosure limiting techniques and causal estimators for discussion's sake. Further evaluation of disclosure limiting transformations and causal estimators would be valuable. For example, the inverse propensity score weighted estimator (IPSW) is an estimator of the PATE, which combines experimental and observational data \cite{colnet_causal_2021}. The IPSW estimator requires pooling experimental and observational data, and therefore requires subject-level data. Future work could evaluate the utility of different methods for synthetic data generation of observational data for use in the IPSW estimator. The results also point to more work to be done to generate differentially private Gram matrices with higher utility for OLS estimation and inference and to be used in the data integration approach. 

Integrating observational and experimental data in treatment effect estimation is a powerful and exciting direction in causal inference. There is a lost opportunity when RCT analysts cannot access relevant observational data. On the other hand, individuals who provide their data have the right to control their information and expect that sensitive information will not be released to the public. In practice, choosing a tolerable disclosure risk, balanced with data utility, is a policy decision. In this paper, we present information to inform such a decision, illustrating the privacy-utility trade-off for different data privacy techniques when integrating private auxiliary data with experimental data for causal inference. 
 
\label{sec:disc}

\section{Code and Data}

All code to replicate the simulations is available at \url{https://github.com/manncz/exp-obs-priv}.

\section{Acknowledgements}
The research reported here was supported by the Institute of Education
Sciences, U.S. Department of Education, through Grant R305D210031 to the
University of Michigan. The opinions expressed are those of the authors and do not represent views of the Institute or the U.S. Department of Education nor other funders. Charlotte Z. Mann was additionally supported by the National Science Foundation RTG grant DMS-1646108. 


\bibliographystyle{vancouver}
\bibliography{ref}

\newpage
\appendix

\clearpage
\pagenumbering{arabic}  
\setcounter{page}{1}

\section{Sensitivity Calculations for Differentially Private Gram Matrix Algorithm}\label{appx:dpg-matrix}

Take an $n\times p$ dataset $\bm X$. Let $\bm x$ and $\bm y$ be some columns of $X$, and $x_i$ is the $i$th element of the vector $\bm x$. We are interested in calculating the $\ell_2$ sensitivity:
$$\Delta f =  \max\limits_{\bm X,\bm X'}||f(\bm X') - f(\bm X)||_2$$
where $d(\bm X,\bm X') = 1$ or in other words, they differ by one observation. We will assume that $d(\bm X,\bm X') = 1$ indicates that $\bm X'$ is a subset of $\bm X$, with one observation removed. For simplicity, and without loss of generality, let's just say that observation $i = 1$ is the one that is removed in the calculations below.

\subsection{Mean Sensitivity}

We will look at the sensitivity for the empirical mean of one column. We can also think of this as the sensitivity to the sum of the values of one column, removing one observation. Assume that $\bm x$ is bounded below by $b$ and above by $B$.

\begin{align*}
n \Delta f& = \max\limits_{\bm X,\bm X'} |\sum_{i=1}^nx_i - \sum_{i=2}^nx_i| \\
&=  \max\limits_{\bm X,\bm X'} |x_1|\\
   &= \max(|B|,|b|)
\end{align*}

So for the empirical mean, $\Delta f = \frac{1}{n}\max(|B|,|b|)$.

\subsection{Second (Non-Central) Moment Sensitivity}

We will look at the sensitivity for the empirical expectation of the product of $\bm x$ and $\bm y$, defined by $\frac{1}{n}\sum_{i=1}^nx_iy_i$. Assume that $\bm x$ is bounded below by $b_x$ and above by $B_x$ and $\bm x$ is bounded below by $b_y$ and above by $B_y$. Then,

 \begin{align*}
     n \Delta f &=  \max\limits_{\bm X,\bm X'} \Big |\sum_{i=1}^nx_iy_i - \sum_{i=2}^nx_iy_i \Big|\\
     &= \max\limits_{\bm X,\bm X'} \Big | x_1y_1\Big|\\
     &= \max(|B_x|,|b_x|)\max(|B_y|,|b_y|)
 \end{align*}

\section{Design-Based Covariate Adjustment with Auxiliary Data}\label{appx:reloop}

Using a prediction of the outcome as a covariate in the regression estimator is effective to reduce variance in experimental estimates when the model is correctly specified and there is no covariate shift between the RCT and auxiliary study. In practice, neither of these assumptions may be true. Design-based methods for covariate adjustment do not require modeling assumptions on top of those that are typically employed with design-based analysis of randomized experiments. The covariate adjusted estimator proposed by \cite{gagnon-bartsch_precise_2023} is robust to model mis-specification and accounts for possible covariate shift between the RCT and auxiliary study.

\cite{gagnon-bartsch_precise_2023} is part of the literature proposing residualizing the outcomes in the IPW estimator with some function of the covariates, as in the augmented IPW estimator \cite{robins_estimation_1994, scharfstein1999rejoinder, robins2000robust, aronow_class_2013,sales_rebar_2018,wu_loop_2018}. In general, such adjusted estimators take the form:$$ \taipw = \frac{1}{n}\Big[\sum_{i \in \mathcal{R}}T_i\frac{Y_i - \hat{f}(\bxi)}{\pi} - (1-T_i)\frac{Y_i - \hat{f}(\bxi)}{1-\pi}\Big].$$

Authors have proposed different functions $f(\cdot)$ for adjustment. \cite{aronow_class_2013} remain agnostic to an exact choice for $f(\cdot)$, but note that the choice of  $f(\cdot)$ could impact efficiency. \cite{sales_rebar_2018} use $f(\cdot) = y_i^c$ (so $\hat{f}(\bxi)$ estimates the control potential outcome for subject $i$). \cite{wu_loop_2018} propose $f(\cdot)= m_i = \pi y_i^t + (1-\pi)y_i^c$. $m_i$ is not observed, but if $\hat{f}(\bxi)= m_i$, $\taipw = \tau^{SATE}$, minimizing the estimator's variance. Therefore, the better $\hat{f}(\bxi)$ estimates $m_i$, the more precise $\taipw$ will be. 

Using auxiliary, observational data to estimate $f(\cdot)$ can further improve precision because the auxiliary data likely has a much larger sample size than the randomized experiment. \cite{aronow_class_2013,sales_rebar_2018} suggest using auxiliary data directly to estimate $f(\cdot)$, for example, letting $\hat{f}(\cdot)$ be a regression model fit on the auxiliary data. \cite{gagnon-bartsch_precise_2023} instead suggest using predictions of the outcome based on a regression model fit on the auxiliary data as a covariate to estimate $m_i$ in \cite{wu_loop_2018}. 

\begin{figure}
    \centering
    \include{figures/re-design-appx}
    \caption{Simulated relative efficiency of $\tadj(\cdot)$ estimator with different covariates compared with the $\tregrct$ estimator (using RCT covariates only). The relative efficiency, or sample size multiplier, is calculated as $\tregrct/\hat{\tau})$, where  $\hat{\tau}$ is the given estimator. Values larger than 1 mean that the estimator is more efficient. $\yhato(\cdot)$ is the vector of predictions for the RCT sample from a model fit on the auxiliary data using a certain release of the auxiliary data. $\epsilon$ is the privacy budget for DP transformations. Error bars represent two simulation standard errors.}
    \label{fig:re-design}
\end{figure}

Like the regression estimator discussed in Section~\ref{sec:prec_est}, \cite{gagnon-bartsch_precise_2023} only requires a model of the outcome of interest fit on the auxiliary data. We denote the specific variation of the adjusted IPW estimator in \cite{gagnon-bartsch_precise_2023} as $\tadj(\cdot)$. We run the same simulations as in Section~\ref{sec:sim-var}, using $\{\hat{Y}_i(\cdot), \bxi\}$ as covariates in the ensemble approach of  $\tadj(\cdot)$ instead of the regression estimator. The ensemble approach interpolates between a model using the RCT covariates, and a model using only the auxiliary prediction, so it is  robust to covariate shift. We implement the estimator using the \texttt{loop.estimator} package in \texttt{R} \cite{wu_loopestimator_2022} (now replaced by the \texttt{dRCT} package \citep{drct_2023} since the time of original submission). Figure~\ref{fig:re-design} shows that the adjusted IPW estimator $\tadj(\cdot)$ performs similarly to the regression estimator in this setting. However, $\tadj(\cdot)$ is better at guarding against super noisy auxiliary predictions, and always is more efficient than the difference-in-means estimator, even when there are small privacy budgets and $p = 20$ or $p = 50$.

\end{document}